\documentstyle[twocolumn,aps,epsfig]{revtex}

\begin{document}

\title{
\vspace{-1.05cm}
\begin{flushleft}
{\normalsize DESY 97-252} \hfill\\
\vspace{-0.15cm}
{\normalsize HUB-EP-97/88} \hfill\\
\vspace{-0.15cm}
{\normalsize December 1997} \hfill\\ 
\vspace{-0.15cm}
\end{flushleft}
Is there a Landau Pole Problem in QED?}

\author{M.~G\"ockeler$^a$,
        R.~Horsley$^b$,
        V.~Linke$^c$,
        P.~Rakow$^d$, 
        G.~Schierholz$^{d,e}$
        and H.~St\"uben$^f$ \\[0.6em]}

\address{$^a$ Institut f\"ur Theoretische Physik, Universit\"at Regensburg,
           D-93040 Regensburg, Germany\\
      $^b$ Institut f\"ur Physik, Humboldt-Universit\"at zu Berlin,
      D-10115 Berlin, Germany\\
      $^c$ Institut f\"ur Theoretische Physik, Freie Universit\"at Berlin,
      D-14195 Berlin, Germany\\
      $^d$ Deutsches Elektronen-Synchrotron DESY, HLRZ and Institut f\"ur
           Hochenergiephysik, D-15735 Zeuthen, Germany\\
      $^e$ Deutsches Elektronen-Synchrotron DESY,
      D-22603 Hamburg, Germany\\
      $^f$ Konrad-Zuse-Zentrum f\"ur Informationstechnik Berlin,
      D-14195 Berlin, Germany\\}
%


\maketitle

\begin{abstract}
We investigate a lattice version of QED by
numerical simulations. For the renormalized charge and mass we find results
which are consistent with the renormalized charge vanishing in the continuum
limit. A detailed study of the relation between bare and renormalized 
quantities reveals that the Landau pole lies in a region of parameter
space which is made inaccessible by spontaneous chiral symmetry breaking.
\end{abstract}

\section{Introduction}

QED is the best tested of all quantum field theories.  But all its
success is in the context of perturbation theory. It has long
been known that there are potential problems in the foundations of
the theory due to the existence of the so-called Landau pole~\cite{Landau}.
In the leading logarithmic calculation one finds 
\begin{equation}
\frac{1}{e_R^2} - \frac{1}{e^2} = \beta_1 \ln\frac{\Lambda}{m_R},\;
\beta_1 = \frac{N_f}{6 \pi^2},
\label{e}
\end{equation}
where $e$ ($e_R$) is the bare (renormalized) charge, $m_R$ is the
renormalized fermion mass, $N_f$ is the number of flavors and
$\Lambda$ is the ultra-violet cutoff.  When one attempts to send
the cutoff to infinity while keeping $e_R$ fixed, one finds that $e$
diverges at
\begin{equation}
\Lambda = \Lambda_L \equiv m_R \mbox{e}^{\frac{1}{\beta_1 e_R^2}},
\end{equation}
the location of the Landau pole.  The
problem can also be seen by looking at the gauge invariant part of the
photon propagator
\begin{equation} 
\frac{D(k)}{k^2} = \frac{1}{k^2\,[1 - (\beta_1/2) \ln(k^2/m_R^2)]}\,,
\label{D}
\end{equation}
which has a ghost pole at $k^2 = \Lambda_L^2$. This would mean that the
entire theory is only applicable for momenta smaller than $\Lambda_L$.
On the other hand, when one keeps $e$ fixed and sends the cut-off to 
infinity, the renormalized charge goes to zero, meaning that the theory
is trivial. The situation in two-loop perturbation theory is much the 
same. The (renormalized) $\beta$ function
\begin{equation}
\beta_R \equiv \left. m_R \frac{\partial e_R^2}{\partial m_R}\right|_{e^2} =
\beta_1 e_R^4 + \beta_2 e_R^6 + \cdots
\end{equation}
remains positive for all $e_R^2$, and the Landau pole is displaced to
lower values 
\begin{equation}
\Lambda_L = m_R \mbox{e}^{\frac{1}{\beta_1 e_R^2}}
\left(\frac{\beta_2 e_R^2}{\beta_1 + 
\beta_2 e_R^2}\right)^{\frac{\beta_2}{\beta_1^2}}\,.
\end{equation}

QED is not the only theory with a Landau pole problem. Every theory which
is not asymptotically free suffers from this problem. While 
$\Lambda_L \simeq 10^{227} \, \mbox{GeV}$ if only the electron is considered,
$\Lambda_L \simeq 10^{34} \, \mbox{GeV}$ in the Standard Model.
In the Minimal Supersymmetric Standard Model (MSSM)
$\Lambda_L \simeq 10^{20} \, \mbox{GeV}$, and in the MSSM with four Higgses, 
which offers a solution to the strong CP problem, the Landau pole 
moves down to $\Lambda_L \simeq 10^{17} \, \mbox{GeV}$~\cite{Yndurain}.
Thus the Landau pole is by no means academic.

To find a solution to this problem, one must consider a
non-perturbative formulation of QED.  Thus it is natural to
investigate the problem on the lattice. On the lattice the inverse lattice
spacing takes over the role of the ultra-violet cut-off, 
$a^{-1} \sim \Lambda$. Early calculations have shown
that the non-compact formulation of the theory using staggered
fermions undergoes a second order chiral phase transition at strong
coupling~\cite{Kogut,DESY}:
\begin{figure}[b,h]
\vspace*{-7.0cm}
\begin{centering}
\epsfig{figure=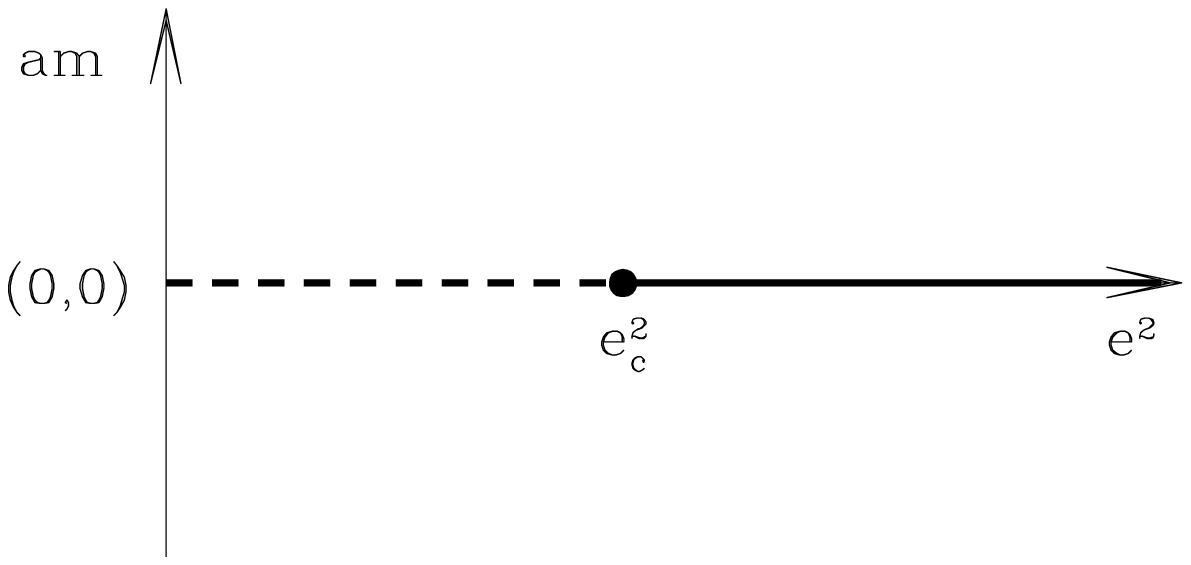,height=11cm,width=10cm}
\end{centering}
\vspace*{-2.0cm}
\end{figure}
\noindent
The solid line $am = 0$, $m$ being the bare mass, $e^2 > e_c^2$ is a 
line of first order chiral phase transitions, where 
$a m_R$, $a^3 \langle\bar{\psi}\psi\rangle \neq 0$, even though the bare
mass is zero. The dashed line $am = 0$, $e_c^2 > e^2$ is a line of second 
order phase transitions on which $a m_R$, 
$a^3 \langle\bar{\psi}\psi\rangle  = 0$. A meaningful continuum limit can be 
taken at the tricritical point 
$am = 0$, $e^2 = e_c^2$, because here we can
take $a$ to zero while keeping $m_R$ fixed. 

To understand the continuum limit of the theory, we need to know the
renormalized charge as a function of the cut-off in the 
critical region. We have recently computed the chiral condensate on large 
lattices~\cite{eos}. In this letter we compute $a m_R$ 
and $e_R$  
in order to understand the fate of the 
Landau pole in QED. 

\begin{figure}[t,h]
\begin{centering}
\epsfig{figure=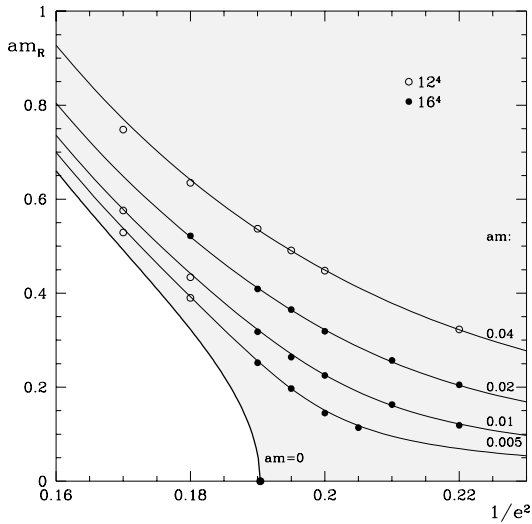,height=7.5cm,width=7.5cm}
\caption{The renormalized mass against the bare coupling on $12^4$ and
$16^4$ lattices.} 
\end{centering}
\end{figure}

\section{Fermion Mass}

We obtain the renormalized mass from the fermion propagator as outlined in
Ref.~\cite{big}. We are using staggered fermions which in the continuum limit
correspond to $N_f = 4$ flavors of dynamical Dirac fermions. The results 
are shown in Fig.~1.

Fitting and extrapolating this data it greatly helps that the chiral condensate
$\sigma = a^3 \langle\bar{\psi}\psi\rangle$ is a function of $a m_R$ 
only~\cite{big}. In Fig.~2 we plot $\sigma$ as a function of $a m_R$. We 
observe that $\sigma$ is well described by the polynomial
\begin{eqnarray}
\sigma &=& 0.6197 a m_R - 0.321 (a m_R)^3 + 0.169 (a m_R)^5 \nonumber\\
 & &  - 0.040 (a m_R)^7,
\label{pol}
\end{eqnarray}
where the first coefficient is given by the one-loop result. This helps because
we have already found an equation of state (EOS) that describes the $\sigma$ 
data~\cite{eos}. Combining the EOS with the 
polynomial (\ref{pol}) gives the curves shown in Fig.~1 and the extrapolation 
to $am = 0$. (Here we have used fit 1 of Ref.~\cite{eos}. Our results 
do not change qualitatively if we use any of the other fits 
described there.)
For $1/e^2 < 1/e_c^2$ chiral symmetry is broken, and even at 
$am = 0$ the renormalized mass is non-zero. This means there is an excluded
region shown in white in Fig.~1 (the accessible region being shown in gray).

\section{Renormalized Charge}

The renormalized charge is obtained from the residue of the photon propagator,
$e_R^2 = Z_3 \, e^2$ and 
$Z_3 = \lim_{k\rightarrow 0} \lim_{V\rightarrow \infty} D(k)$. We can compute
$D(k)$ on the lattice, but not at $k = 0$. The smallest momentum 
that we 
can reach is $2\pi/aL$, \hfill where $L$ is the lattice size ($L = 16$ \\

\begin{figure}[th]
\begin{centering}
\epsfig{figure=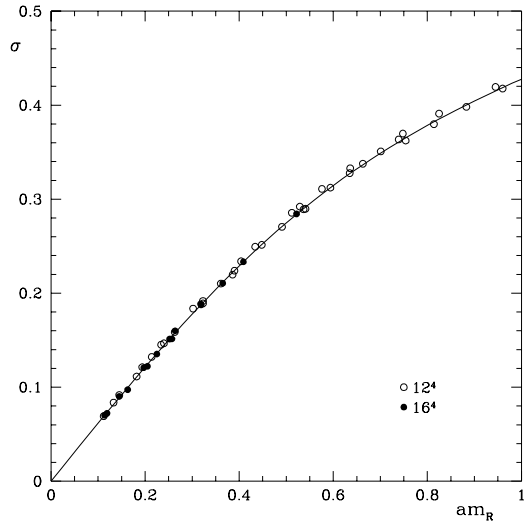,height=7.5cm,width=7.5cm}
\caption{The chiral condensate against the renormalized mass on $12^4$ and
$16^4$ lattices.} 
\end{centering}
\end{figure}

\noindent
and $12$ in our case). To extrapolate to $k = 0$ we need to make a fit to 
the photon propagator. The $k$ dependence of the photon propagator is given by
\begin{equation}
\frac{1}{e^2 D(k)} - \frac{1}{e^2} = - \Pi(k,m_R,L)\,,
\end{equation}
where $\Pi$ is the polarization function. In the infinite volume limit we
then have
\begin{equation}
\frac{1}{e_R^2} - \frac{1}{e^2} = - \Pi(0,m_R,\infty)\,.
\label{er}
\end{equation}
We have already seen that the non-perturbative $\Pi$ is actually very close
to the result of one-loop renormalized perturbation theory~\cite{big}. So
it is reasonable to make an ansatz which is inspired by renormalization
group improved two-loop perturbation theory. In Ref.~\cite{Shirkov} it is 
shown that to next-to-leading logarithmic order the polarization function can be written 
\begin{equation}
\Pi = U - \frac{V}{U} \ln(1-e^2U)\,,
\end{equation}
where $U$ is the one-loop perturbative result, and $V$ the two-loop one. The
lattice result for $U$ is known~\cite{big}. For $V$ we make the ansatz
\begin{equation}
V = v_0 + v_1 U\,.
\end{equation}
This is motivated by the small $k^2$ and $m_R^2$ limits. For
$a^2 m_R^2 \ll a^2 k^2 \ll 1$ we should have 
$V \simeq (\beta_2/2) \ln a^2 k^2$,
and for $a^2 k^2 \ll a^2 m_R^2 \ll 1$ we should have
$V \simeq (\beta_2/2) \ln a^2 m_R^2$. The one-loop result $U$ has these 
properties. We fit this ansatz to a total of 52 photon propagators on 
$16^4$ and $12^4$ lattices for various values of $am$, 
$e^2$ in the range
$0.005 \leq am \leq 0.16$, $0.16 \leq 1/e^2 \leq 0.22$ 
close to the 
critical point at $1/e_c^2 = 0.19040(9)$~\cite{eos}. \hfill A plot for \\

\begin{figure}[t,h]
\begin{centering}
\epsfig{figure=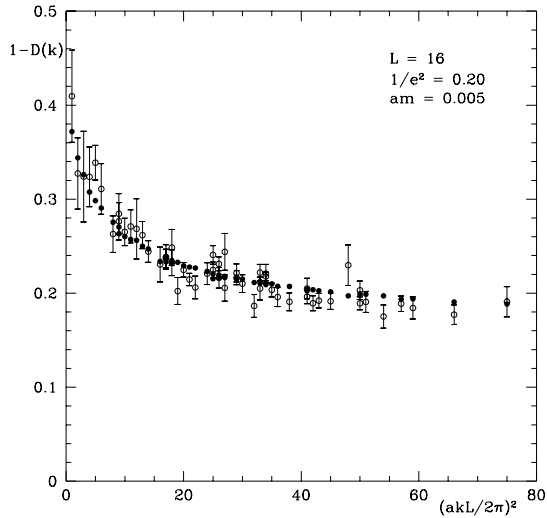,height=7.5cm,width=7.5cm}
\caption{The residue of the photon propagator against the momentum for
$1/e^2 = 0.20, am = 0.005$ on the $16^4$ lattice. The open 
symbols are the data, the solid symbols are the fit.} 
\end{centering}
\end{figure}

\noindent
one particular parameter
set is shown in Fig.~3. For the fit parameters we obtain $v_0=-0.00207(2)$
and $v_1=-0.0328(7)$, giving $\chi^2/\mbox{d.o.f.} = 1.7$. Two-loop continuum
perturbation theory would give $v_1\equiv\beta_2/\beta_1 =
3/16\pi^2 = 0.0190$. In Fig.~4 we show the resulting $\beta$ function
for $e^2 = e_c^2$. We compare this with the one-loop result. We see that the 
$\beta$ function is a little smaller 
than the one-loop value and is positive. 
In particular this means that there is no
ultra-violet stable zero in the $\beta$ function out to $e_R^2 = e_c^2$,
the maximal value $e_R^2$ can take because $Z_3 \leq 1$~\cite{BD,Luscher}.
As $am_R \rightarrow \infty$ fermion loops are suppressed and 
$e_R^2 \rightarrow e^2$, so that the $\beta$ function vanishes. But this is
of course not an interesting zero of the $\beta$ function.

\begin{figure}[b,h]
\begin{centering}
\epsfig{figure=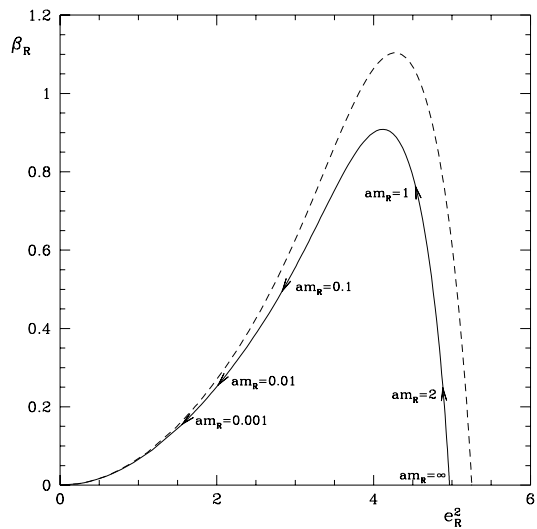,height=7.5cm,width=7.5cm}
\caption{The $\beta$ function against the renormalized charge. The solid curve
is our result, the dashed curve is the lattice one-loop result.} 
\end{centering}
\end{figure}
\begin{figure}[t,h]
\begin{centering}
\epsfig{figure=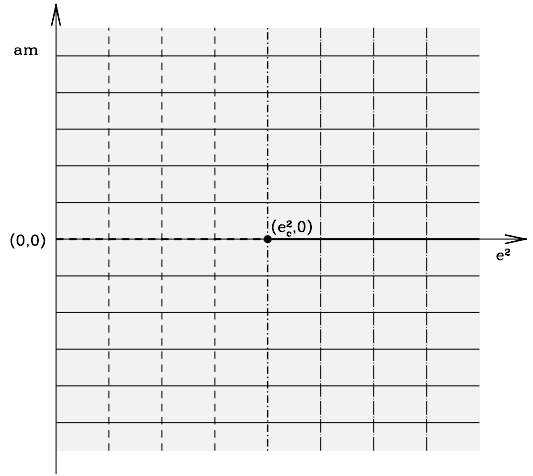,height=7.5cm,width=7.5cm}
\epsfig{figure=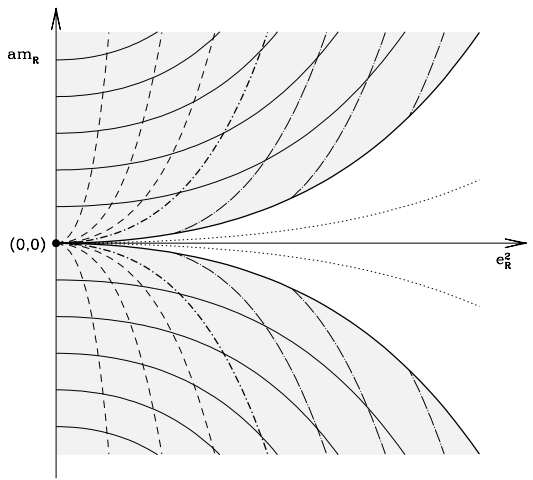,height=7.5cm,width=7.5cm}
\caption{A sketch of the mapping from the bare parameter plane (top) to the
renormalized parameter plane (bottom).} 
\end{centering}
\end{figure}

\section{The Landau Pole} 

Having calculated the renormalized mass and charge, we are now able
to discuss the mapping from the bare parameters $am$, $e$ to the
renormalized parameters $am_R$, $e_R$. Qualitatively this is displayed in
Fig.~5. One can choose any $e^2 \geq 0$, $am$ shown in the top part of 
the figure by the gray
region. This is then mapped onto the corresponding gray region in the bottom
part of the figure.
The line $am = 0$ from $e^2 = 0$ to 
$e^2 = e_c^2$ is mapped onto the point
$am_R = e_R^2 = 0$. For $e^2 > e_c^2$ we have already seen in Fig.~1 that 
$am_R > 0$, even when $am = 0$, because of chiral symmetry breaking.
Thus the line $am = 0$ from $e^2 = e_c^2$ to $e^2 = \infty$ is mapped onto the
border line of the gray regions, and the white area is inaccessible for
any combination of bare parameters. From this figure we evidently have 
triviality. Removing 
\begin{figure}[t,h]
\begin{centering}
\epsfig{figure=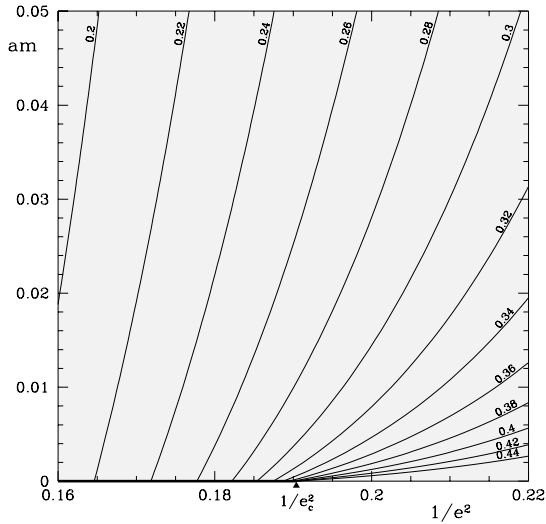,height=7.5cm,width=7.5cm}
\epsfig{figure=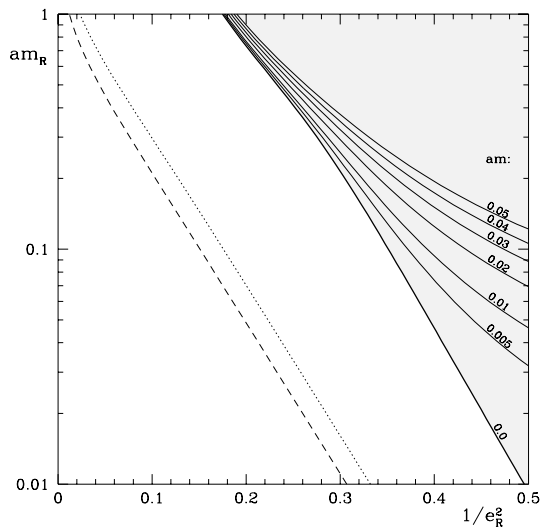,height=7.5cm,width=7.5cm}
\caption{The (quantitative) mapping of the bare parameter plane (top) to
the renormalized parameter plane (bottom).} 
\end{centering}
\end{figure}
\noindent
the cut-off ($a \rightarrow 0$) is only possible at
$e_R^2 = 0$.
At any 
finite value of $e_R^2$ there is a minimal possible value for $|am_R|$, 
namely the boundary of the gray region. The position of the 
Landau pole is sketched by the dotted line.

We now turn to the quantitative analysis of the problem. In Fig.~6 we plot 
again the bare and renormalized planes, this time using $1/e^2$ and 
$1/e_R^2$, respectively, as the 
horizontal variables because this displays the asymptotic behavior best. The 
curves are lines of constant $1/e_R^2$ (top part) and lines of constant $am$
(bottom part), respectively. All lines of constant $e_R^2$ end on the first 
order phase transition line, and only 
the line $e_R^2 = 0$ goes into the 
critical point. This is another expression of triviality of the theory.
In the bottom part of the figure the gray region is again the allowed region, 
and the white 
region is inaccessible. The border line is the line 
$am = 0$. From eq.~(\ref{er}) we find the Landau pole by setting the bare 
charge to infinity. This gives the dotted line. We see that it 
completely lies in the inaccessible region. It runs roughly parallel to the
border line $am = 0$.

We also want to be sure that the photon propagator has no ghost pole for any
$k^2$. Beyond leading logarithmic order the extra pole in the photon 
propagator and the divergence in $e^2$ need not appear at the same place. The
ghost pole position (in the infinite volume) is given by $1/k^2 D(k) = 0$.
It has a solution if
\begin{equation}
\frac{1}{e_R^2} < \max_k \Pi(k,m_R,\infty) - \Pi(0,m_R,\infty).
\end{equation}
The solution is given by the dashed line in the bottom part of Fig.~6. This 
lies close to
the Landau pole, even deeper in the inaccessible region.

\section{Conclusion}

From Figs.~5,~6 we see that the triviality of QED is intimately connected
with chiral symmetry breaking. Any attempt to remove the cut-off is always
thwarted by the dynamically generated fermion mass. In particular this
means that spinor QED does not exist as an interacting theory, similar to what
Coleman and Weinberg~\cite{CW} found for scalar QED.

We have also seen that chiral symmetry breaking allows QED to escape the
Landau pole problem. While the bare parameters of the theory can take any 
values, the renormalized parameters are restricted. The Landau pole and
ghost problem only occur deep in the inaccessible $e_R^2$, $am_R$ region.
Chiral symmetry breaking is always strong enough to push the Landau pole above
the cut-off.

\end{document}